\documentclass[pre,twocolumn,showpacs,amsmath,amssymb,superscriptaddress]{revtex4}
\usepackage{graphicx,amstext}
\usepackage{dcolumn}
\usepackage{bm}

\newcommand{\ie}{{\it i.e.\@}}
\newcommand{\al}{{\it et al.\@}}
\newcommand{\bq}{\begin{equation}}
\newcommand{\eq}{\end{equation}}
\begin{document}
\title{Manning free counterions fraction for a rod-like polyion - short DNA fragments in very low salt}

\author{T.~Vuleti\'{c}}
\homepage{http://tvuletic.ifs.hr/}
\email{tvuletic@ifs.hr}
\affiliation{Institut za fiziku, 10000 Zagreb, Croatia} 
\author{S.~Dolanski Babi\'{c}}
\affiliation{Institut za fiziku, 10000 Zagreb, Croatia}
\altaffiliation {Permanent address: Department of physics and biophysics,
Medical School, University of Zagreb, 10000 Zagreb, Croatia.}
\author{D.~Grgi\v{c}in}
\author{D.~Aumiler}
\affiliation{Institut za fiziku, 10000 Zagreb, Croatia}
\author{J.R\"{a}dler}
\affiliation{Ludwig-Maximilians-Universit\"{a}t, Sektion Physik,
Geschwister-Scholl-Platz 1, D-80539 Munich, Germany}
\author{F.~Livolant}
\affiliation{Laboratoire de Physique des Solides, Universit\'{e} Paris Sud -
F-91405 Orsay, France}
\author{S.~Tomi\'{c}}
\affiliation{Institut za fiziku, 10000 Zagreb, Croatia}
\date{\today}

\begin{abstract}

We quantified the Manning free (uncondensed) counterions fraction $\theta$ for 
dilute solutions of rod-like polyions - 150bp DNA fragments, in very low salt 
$<0.05$mM. Conductivity measurements of aqueous DNA solutions in the 
concentration range $0.015\leq c \leq 8$~mM (bp) were complemented by 
fluorescence correlation spectroscopy (FCS) measurements of the DNA polyion 
diffusion coefficient $D_p(c)$.  We observed a crossover in the normalized 
conductivity $\sigma(c)/c$ which nearly halved across $c=0.05-1$ mM range, while 
$D_p(c)$ remained rather constant, as we established by FCS. Analyzing these 
data we extracted $\theta(c)=0.30 - 0.45$, and taking the Manning asymmetry 
field effect on polyelectrolyte conductivity into account we got 
$\theta(c)=0.40-0.60$. We relate the $\theta(c)$ variation to gradual DNA 
denaturation occuring, in the very low salt environment, with the decrease in 
DNA concentration itself. The extremes of the experimental $\theta(c)$ range 
occur towards the highest, above 1 mM and the lowest, below 0.05 mM, DNA 
concentrations, and correspond to the theoretical $\theta$ values for dsDNA  
and ssDNA, respectively. Therefore, we confirmed Manning 
condensation and conductivity models to be valuable in description of dilute solutions of rod-like polyions.

\end{abstract}



\maketitle

\section{Introduction}
\label{sec1}

Most biologically relevant macromolecules (DNA, proteins, polysaccharides) are 
polyelectrolytes with a very distinct behavior compared to neutral polymers or 
simple electrolytes \cite{Polymerbooks,Bloomfield00}. When dissolved in polar 
solvents polyelectrolytes dissociate into a highly charged polyion (a 
macromolecule of extended shape) and many small counterions of low valency. The 
long range nature of the electrostatic interactions and the entropy effects due 
to inhomogeneities in the counterion distributions and to a myriad of polyion 
configurations control their phenomenology.

The strong linear charge of the polyion tends to attract the counterions to its 
immediate vicinity. The condensation occurs for polyions with the Manning 
parameter $u= l_B/b>1$, where $l_B$ is the Bjerrum length, the length at which 
two elementary charges interact in a given solvent with energy equal to the 
thermal energy $kT$, while $b$ is the average distance between the charges on 
the polyion backbone. If there is more than one charge per Bjerrum length, the 
condensation will tend to effectively reduce the linear charge density down to 
1$/l_B$ level. The condensed ions fraction is then equal to $1-1/u$ and the 
free, uncondensed counterions fraction is $\theta=1/u$. The condensation was 
modeled for an infinitely long and thin polyion in pure water, with no added 
salt, which might appear as a rather unrealistic proposition, with no biological 
relevance \cite{Manning69}. 

Counterion condensation is therefore more easily experimentally studied and the 
results theoretically interpreted for a dilute solution of rigid, monodisperse 
polyions, which do not change conformation with concentration. In a dilute 
solution, effectively, the condensed fraction of counterions may be considered 
to be found in a cylindrical cell around the polyion, while the rest may be 
taken to be free inside a larger volume that belongs to a given 
polyion \cite{Deshkovski01}. According to the theory, since the condensed 
counterions are not chemically bound to the polyion, the free and condensed 
counterions exchange between the two concentric regions and only a continuous 
radial counterion distribution \cite{LeBretZimm84} can exist around the polyion. 
In other words, there should be no step in the radial counterion distribution 
which would define the limit of the cylindrical zone, as shown experimentally by 
EPR (electron paramagnetic resonance) \cite{Hinderberger05}.

Besides theoretical works considering two types of ions, experiments also 
attempt to quantify the condensed and free counterion fractions. Since only 
uncondensed, free counterions contribute  to the osmotic pressure of a 
polyelectrolyte \cite{Hansen01,Antypov06}, the measured osmotic pressure of a 
polyelectrolyte solution evaluates the free counterions fraction 
\cite{Auer69,Raspaud00,Baigl05}. The condensed counterions are those that move 
together with the polyion when an electric field is applied, while the free 
counterions would move in the opposite way  due to their opposite charge 
\cite{Bordi04,Netz08,Bordi02}. Thus, the concept of two types of counterions 
gets a physical meaning. 

Thus, the transport techniques may contribute to our knowledge of condensation 
in polyelectrolytes. The techniques range from electrical transport measurements 
like conductometry \cite{Bordi04,Truzzolillo09,Wandrey99,Wandrey00}  and 
capillary electrophoresis \cite{Tirado84,Stellwagen03}, to  diffusion 
measurements by dynamic light scattering 
\cite{Tirado84,Stellwagen03,Mandelkern81,Pecora05} or fluorescence correlation 
spectroscopy \cite{Wilk04,Hess02}. Manning \cite{Manning81,Manning75} has 
proposed a rather comprehensive and convincing conductivity model for 
polyelectrolytes and Bordi \al \cite{Bordi02} worked on including the scaling 
theories by Rubinstein \al \cite{Dobrynin}, in order to separate the influences 
from the polyion (conformation and charges), the counterions and the added salt.  

For a successful quantitative study of Manning condensation by the transport 
experiments one has to use the simplest possible system: a dilute solution of 
monodisperse polyelectrolytes with no added salt. Also, an experimental method 
is needed to separate the influence of the charge and conformation of the 
polyion on the (electrical) transport. Few experimental works met those 
requirements \cite{Auer69}.  In other cases, there was a necessity to introduce 
the model for the conformation of the polyions into the interpretation of the 
conductivity data \cite{Bordi04,Truzzolillo09,Wandrey99,Wandrey00} which hinders 
the quantification of  the Manning free counterions fraction.

Electrical conductivity in the system under study is a product of three separate 
factors characterizing the mobile charge carriers, summed over all charge 
species $i$ in the system: their charge $z_i\mathrm{e}$, their concentration 
$n_i$ and their mobility $\mu_i$ (ratio of carrier velocity and the applied 
electric field).

\begin{equation}
\sigma=\sum_i (|z_i|\mathrm{e}) n_i \mu_i
\label{zeenmju}   
\end{equation}
For simple electrolytes ({\em cf.} \cite{Bordi02,Bordi04}), it is convenient to  
work with molar concentrations $c_i=n_i/N_A$ and equivalent conductivities 
$\lambda_i=F \mu_i$ (Faraday constant $F=\mathrm{e}N_A$ and $N_A$ is Avogadro 
number). The conductivity is a sum of equivalent  conductivities of the ionic 
species present in solution, multiplied by the charge (valence) $z_i$ and 
concentration $c_i$ of the respective ion:

\begin{equation}
\sigma=\sum_i z_i c_i \lambda_i
\label{zecelambda}   
\end{equation}

For polyelectrolytes the expression of Eq.~\ref{zecelambda} is still valid. A 
monodisperse dilute polyelectrolyte with no added salt will contain only two 
ionic species. For one species, the polyion, a large molecule with a relatively 
small concentration $c_p$ and a proportionally large charge $Z_p$, the equivalent 
conductivity $\lambda_p$ is dependent on its size and conformation, thus

\begin{equation}
\sigma= Z_p c_p \lambda_p + z_i c_i \lambda_i
\label{zecelambda_2}   
\end{equation}
Here we note that $c_i$ is the concentration of counterions released from the 
polyelectrolyte upon solvation, and is proportional to the concentration of 
monomers $c$ constituting the polyion. The monomer concentration $c$ 
is related to the polyion concentration $c_p$ via:

\begin{equation} 
c=Nc_p 
\label{cepeN}   
\end{equation}
where $N$ is the polyion degree of the polymerization. Also, the polyion charge 
is related to the monomer charge $z_p$:

\begin{equation}
Z_p=N z_p
\label{zepeN}   
\end{equation}
Due to the electroneutrality of the solution

\begin{equation}
Z_p c_p=z_i c_i=z_pc
\label{Zczc}   
\end{equation}
Thus the conductivity of a polyelectrolyte solution principally depends on the 
concentration of the monomers $c$:

\begin{equation}
\sigma= z_p c (\lambda_p + \lambda_i)
\label{zecelambda_3}   
\end{equation}
Here we remind that the polyion charge is effectively reduced, $Z_p=\theta N 
z_p$ due to the counterion condensation, and also that only the free fraction 
$\theta c_i$ of counterions is considered to take part in electrical 
transport. Therefore:

\begin{equation}
\sigma= \theta z_p c (\lambda_p + \lambda_i)
\label{sigmafzclambda}   
\end{equation}
The polyion conductivity $\lambda_p$, being defined by polyion mobility, 
actually stems from the self-diffusion coefficient of the polyion $D_p$ and its 
charge $Z_p$, according to Einstein's relation for a charged particle:

\begin{equation}
D_p= \frac{kT \mu_p}{{\mathrm e}Z_p}
\label{Einstein}   
\end{equation}
and thus

\begin{equation}
\lambda_p= F Z_p {\mathrm e} \frac{D_p}{kT}
\label{FZDkT}   
\end{equation}
The diffusion coefficient depends on the size and shape of the particle, as well 
as on the viscosity of the solution in which the particle is moving. Inserting 
Eq.\ref{FZDkT} into eq.\ref{sigmafzclambda} we get

\begin{equation}
\sigma= \theta z_p c (F \theta N z_p {\mathrm e} \frac{D_p}{kT} + \lambda_i)
\label{IntroQuad}   
\end{equation}

Consequently, the conductivity of a monodisperse polyelectrolyte without added 
salt is primarily governed by the self-diffusion coefficient $D_p$ of its 
polyion and the free counterion fraction $\theta$.

In order to quantify the effects of the diffusion and electrostatics in a 
polyelectrolyte we used nucleosomal DNA fragments 150 bp (50 nm) long. These are 
expected to be rather rigid and rod-like since the DNA persistence length is 50 
nm \cite{Baumann97}. The dilute-semidilute crossover concentration for these 
fragments is $\approx$2 mM \cite{deGennes76}. The details of material 
preparation and experimental methods are given in Sec. II.  As presented in Sec. 
III, conductivity measurements were complemented by fluorescence correlation 
spectroscopy (FCS) measurements of the DNA polyion self-diffusion coefficient 
$D_p$. Our proposition, discussed in Sec. IV is that the conductivity crossover  
observed in $c=0.05-1$ mM (in basepair) DNA concentration range results from the DNA 
denaturation that induces a concomitant change in the extent of Manning 
condensation. Eventually, we estimate the free counterions fraction $\theta$ and 
compare them with the values predicted by Manning for both ssDNA and 
dsDNA.  

\section{Materials \& Methods}
\label{sec2}
\subsection{Monodisperse DNA}
We will express DNA concentrations as molar concentrations of basepairs (bp) 
(1g/L equals 1.5 mM bp).

Large quantities of practically monodisperse nucleosomal DNA fragments  were 
prepared as described in Sikorav \al \cite{Sikorav94} by enzymatic digestion of 
H1 depleted calf thymus chromatin \cite{StrzeleckaRillJACS87}. This DNA, denoted 
DNA146, contains fragments $150 \pm 10$bp long (50 nm) together with traces of 
300-350 bp fragments that correspond to two nucleosomal DNA fragments connected 
by undigested linker DNA. DNA fragments were precipitated with cold ethanol, 
dried and stored at 4$^o$C. The stock solution was prepared by dissolving 10 mg 
of the Na-DNA pellet in 0.55 mL pure water. A low protein content was verified 
by UV absorption. DNA146 solutions (0.015 - 8mM bp) were prepared by dilution 
with pure water of aliquots from this 27mM mother solution. To check that no 
salt was released from the pellet in addition to the  Na$^+$counterions (2 
Na$^+$ per bp), an aliquot of the pellet was dissolved in 10 mM NaCl, diluted 5 
times with pure water and spin-filtered to the original volume. This procedure 
was repeated 3 times. Another sample was simply dissolved in pure water. The two 
samples had similar conductivities (normalized for concentration).  We concluded 
that any salt that may have been present in the pellet did not raise the 
conductivity more than the equivalent of 0.2 Na$^+$ ions per basepair.

110bp dsDNA was prepared as follows. Two separate oligonucleotides (ssDNA, 
110nt) were purchased (Mycrosynth A.G., CH) \cite{sqnc}. The two sequences were 
complementary and one of them was labeled at one end with a covalently bound Cy5 
fluorophore. The dry complements were dissolved in 10 mM Tris-EDTA (TE) buffer 
with up to 60mM NaCl, mixed and heated to 97$^{o}$C for 15 minutes 
to remove any hairpin loops previously formed and then left to cool down for 
several hours to slowly hybridize and form 110 bp long dsDNA. Hybridization was 
checked to be complete on an agarose gel. To remove any NaCl excess, the 
solution was diluted 4 times with a large volume of 10mM Tris-Cl- buffer and 
then spin-concentrated to the original volume. The procedure was repeated 3 
times. In the resulting solution, (denoted DNA110* with * to indicate the 
fluorescent labeling), the DNA and Cy5 concentrations were respectively 0.5mM 
and 5$\mu$M. 

For FCS measurements, 2 $\mu$L of DNA110* stock were added into 500 $\mu$L 
DNA146 of varying  concentrations (0.0015-8 mM concentration range) to achieve a 
20 nM Cy5 concentration. The 10mM Tris of the DNA110* stock was diluted 250 
times. Therefore, all experiments were performed at very low salt 
($c_{\mathrm{salt}}<0.05$mM). For FCS calibration, Cy5 fluorophore alone was 
diluted in pure water to 20 nM. Since the fluorophore concentration can deviate 
only less than one order of magnitude from this concentration, the amount of 
fluorescently labeled DNA110* was fixed whereas the concentration of DNA146 
spans over several orders of magnitude.

\subsection{Fluorescence correlation spectroscopy}

Fluorescence correlation spectroscopy inherently probes the system under study 
both at single molecule and ensemble levels. FCS observes fluorescence intensity 
fluctuations emitted by fluorescently labeled objects diffusing through a small 
open volume ($< 1$ fL) defined by the profile of the laser beam and the optics, 
objective of the microsope. That is, number fluctuations of the molecules 
entering and leaving the focal volume are registered as fluorescence variation, 
which is then recorded and autocorrelated. Thus following practically single 
molecules we obtain the properties of the ensemble 
\cite{RiglerBiosci90,SchwilleBPC97}. We have used a commercially availabe Zeiss 
ConfoCor II FCS instrument, where the measurement volume was defined by a Zeiss 
Plan-NeoFluar 100x/NA1.3 water immersion objective, epi-illumination was by 
He-Ne 632.8 nm 5mW laser, for excitation of Cy5 fluorophore. Measurements were 
performed at $25^{o}$C, the ambient temperature of the temperature stabilized 
clean-room. Zeiss proprietary software was used for autocorrelation function 
calculation and extraction of diffusion times by non-linear least squares 
fitting \cite{ZeissManual}.  The physical principles of such an experimental 
set-up and theoretical background of FCS have been described elswhere 
\cite{SchwilleBPC97,Mangenot03}. The manner used to obtain the self-diffusion 
coefficient of the molecule under study, in our case 110 bp Cy5 labelled dsDNA, 
is presented in brief in the following. The instrument directly measures 
fluorescence intensity for {\em e.g.} 30 seconds. The autocorrelation function 
$G(\tau_c)$ is calculated for the intensity trace, with the correlation time 
$\tau_c$ as the variable.  The fluorescence intensity autocorrelation function, 
$G(\tau_c)$, is fitted with a diffusion time, $\tau$. This FCS diffusion time 
relates to the characteristic time for fluorescent  particle to diffuse through 
the focal volume.  Autocorrelation function  decays exponentially and is fitted 
to

\begin{equation}
G(\tau_c) = \frac{1}{N_f} \cdot \frac{1} {1 +\frac{\tau_c}{\tau}} \frac{1}{\sqrt{(1 + (\frac{w_0}{z_0})^2\frac{\tau_c}{\tau})}} (1 + \frac{T}{1-T}exp(-\frac{\tau_c}{\tau_T}))
\label{FCSautocorr}
\end{equation}

Here $N_f$ is average number of fluorescent molecules in the confocal detection 
volume. The transition of the Cy5 fluorophore to the first excited triplet state 
and a relatively slow relaxation to ground state influence the observed 
autocorrelation curve. Thus, $T$, average fraction of fluorophores in the 
triplet state, and $\tau_ T$, lifetime of the triplet state of the fluorophore 
are taken into account when fitting. Another fit parameter is $z_0/w_0$, the 
structure parameter, \ie the ratio of the axial and radial extension of the 
focal volume. The structure parameter $z_0/w_0\approx10$ is obtained from fits 
to autocorrelation curves measured for Cy5 molecules in pure water solution. 
Then it is kept as a fixed parameter when $\tau$ is later being extracted for 
DNA110*. The self-diffusion coefficient $D_p$ of any particle is easily obtained 
from its FCS diffusion time $\tau$ as these are inversely proportional. Since 
the diffusion coefficient of Cy5 is known, $D_{Cy5}=3.16 \cdot 10^{-10}$ m$^2$/s 
\cite{ZeissManual} and the diffusion time $\tau_{Cy5}$ we found to be about 50 
$\mu$s, this provides means for conversion of the diffusion times $\tau$ into 
$D_p$:
\begin{equation}
D_p=D_{Cy5}\frac{\tau_{Cy5}}{\tau}
\label{DDCy5}	
\end{equation}

\subsection{Conductometry}

Dielectric spectroscopy in the range 100Hz-110MHz was performed with Agilent 
4294A impedance analyzer. All the measurements were performed at $25^{o}$C. 
Conductometry data was extracted from these spectra. Conductivity was calculated 
from conductance at 100 kHz and capacitance was read at 10 MHz. Conductivity at 
100 kHz shows a minimal influence from the electrode polarization effects, as 
well as from the conductivity chamber resonance at 100 MHz. Basically, one has 
to measure a spectrum  \cite{TomicEPL08}, to be able to confidently extract 
conductivity values. Only in this manner, the obtained conductivity may be 
regarded as dc conductivity, the conductivity related to currents of freely 
mobile charges (polyions and free counterions) and not due to 
polarization currents. We emphasize that all the  conductivities of 
polyelectrolytes have been deducted for 1.5 $\mu$S/cm, the conductivity of the 
solvent \cite{TomicPRE07}, \ie pure water (Milli-Q, Millipore). This residual 
conductivity is due to the ambient CO2 dissolved in pure water. In this manner, 
pure water solutions may be regarded as very low salt solutions, 
$c_{\mathrm{salt}}<0.01$mM, and we labeled them appropriately. The pH of pure 
water exposed to air is about 5.5, however this is unbuffered. The capacitance 
at 10MHz serves as a check of the sample volume for our experimental setup 
\cite{VuleticPRE10}. At this high frequency the contribution to the capacitance 
comes from the dielectric constant of pure water, and not from the solutes. Thus 
all the samples should have the same capacitance if they have the same volume.

\section{Results}
\label{sec3}
 
\subsection{Electrical transport}

\begin{figure} 
\resizebox{0.46\textwidth}{!}{\includegraphics*{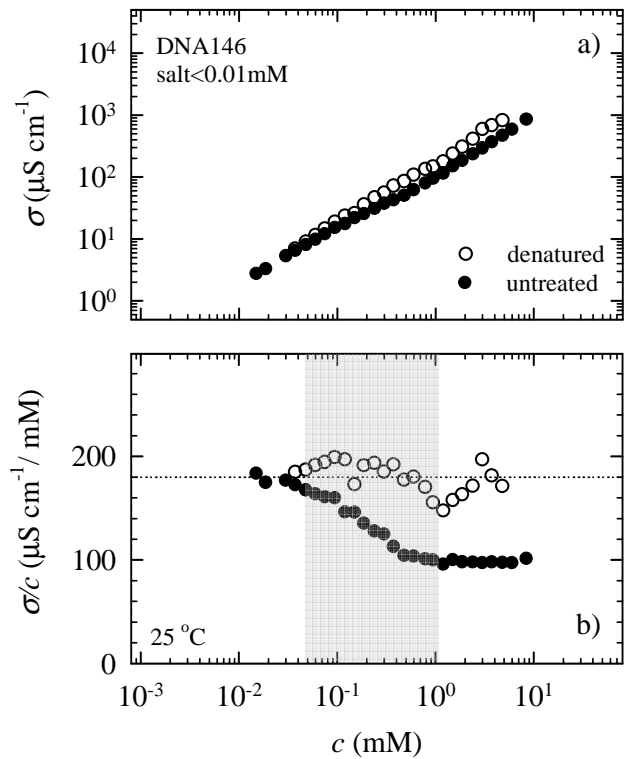}} 
\caption{ 
Conductometry data for DNA146 solutions before (black circles) and after 
denaturation at 97$^{o}$C (open circles). (a) DNA146 solution conductivity versus DNA 
basepair concentration. (b) conductivity normalized by concentration versus DNA 
basepair concentration. The dotted line shows the average value for denatured 
samples, 180 $\mu$Scm$^{-1}/$mM. The values for untreated samples at the lowest 
concentrations also approach this value. Shaded rectangle denotes the crossover 
concentration region.  Measurements were performed at 25$^{o}$C. 
} 
\label{normcond} 
\end{figure} 

We present the dc conductivity data of $0.015-8$ mM DNA146 solutions. 
Experiments were performed at 25$^{o}$C in the absence of added salt 
(concentration of Na+ or Tris ions $<0.05$mM) on the untreated DNA solution and after 
DNA denaturation. Fig.\ref{normcond}(a) emphasizes a general power-law 
dependence of polyelectrolyte conductivity on monomer concentration (see 
Eq.\ref{sigmafzclambda}). However, a slight S-shaped bending may be noted in 
$\sigma(c)$ for the untreated sample (black circles). After 20 min at a 
temperature of 97$^{o}$C, followed by a quenching to 4$^{o}$C for a minute, the 
conductometry was performed at 25$^{o}$C. For these, denatured samples (open 
circles) the power-law is apparently better defined. 

If we normalize the conductivity with concentration, then data may be presented 
in a physically more relevant form, Fig.\ref{normcond}(b). That is, normalized 
conductivity concentration dependence is directly related  to the behavior of 
the molar conductivities of ${{\mathrm {Na}}^+}$ counterions 
$\lambda_i=\lambda_{{\mathrm {Na}}^+}$ and DNA polyions $\lambda_p$:
\begin{equation} 
\frac{\sigma}{c}=2\theta(\lambda_{{\mathrm {Na}}^+}+\lambda_p) 
\label{normcondEq}	
\end{equation} 
Here, the factor 2 stands for DNA monomer charge (valence)$z_p=2$  (see 
Eq.\ref{sigmafzclambda}). We deem that the normalized conductivity (open 
circles) of the denatured samples is, within the data scatter, constant, with a 
value of 180 $\mu$S/cm. The data for denatured samples show a rather high 
scatter, which we ascribe to the denaturation procedure. Either denaturation did 
not proceed to the full extent for all the samples or some of the DNA renatured 
in hairpins during quenching \cite{Montrichok03} and this introduced a 
conductivity variation. Gradual renaturation, after quenching, and during the 
measurement at 25$^{o}$C was not an issue, as the samples held in our 
conductivity chamber showed a stable conductivity for at least an hour, and the 
measurement itself lasted for only 2 minutes.  

Contrary to the denatured DNA, the normalized conductivity of the untreated 
DNA146 samples shows a crossover in the 0.05-1 mM concentration range (the 
crossover region is denoted by a shaded rectangle). Above 1 mM it attains a 
constant value of 100  $\mu$Scm$^{-1}/$mM, while at the lowest concentration it 
approaches the 180  $\mu$Scm$^{-1}/$mM value for the heat treated, denatured 
DNA146 samples. This conductivity crossover has not, to our knowledge, been 
reported previously, for any DNA sample.

\subsection{Polyion diffusion}

\begin{figure}
\resizebox{0.46\textwidth}{!}{\includegraphics*{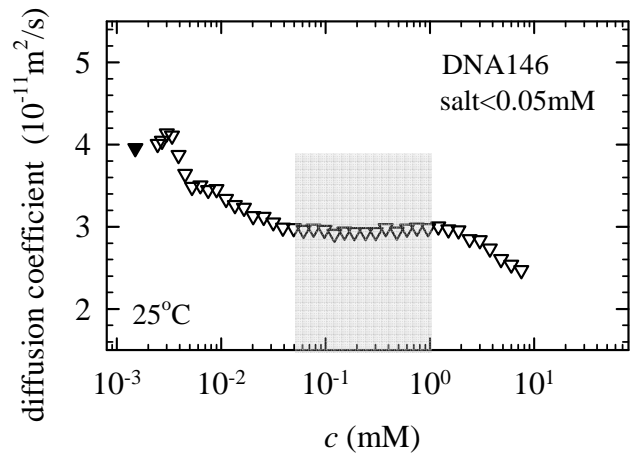}}
\caption{
Diffusion coefficient $D^{exp}_{146}(c)$ for DNA146 polyion, obtained by  
fluorescence correlation spectroscopy (FCS) is shown versus DNA basepair 
concentration.  Shaded rectangle 
denotes the crossover concentration region identified from conductivity 
measurements. Black triangle denotes diffusion coefficient $D^{ss}$ derived for 
146 bp ssDNA.
}
\label{difftime}
\end{figure}

We had to check whether the observed conductivity crossover relates to a change 
in DNA146 conformation due to DNA denaturation expected in the very low salt 
environment  \cite{Record75}. Therefore, we had to obtain the concentration 
dependence of the self-diffusion coefficient $D_p$ for the DNA146 polyion for 
the concentration range studied by conductometry. However, the  FCS diffusion 
times $\tau$ were measured for fluorescently labeled DNA110* polyion diffusing 
freely along the DNA146, but not for DNA146 itself.The labeled DNA is somewhat 
shorter than the bulk of DNA in the sample solution.  Thus diffusion 
coefficients $D^{exp}_{110*}(c)$ for DNA110* that may be derived according to 
Eq.\ref{DDCy5} had to be extrapolated to obtain $D^{exp}_{146}(c)$ values for 
DNA146. That is, DNA110* and DNA 146, 38 and 50 nm long, respectively, have 
lengths comparable to the dsDNA persistence length $L_p=50$ nm \cite{Baumann97}. 
Thus, an extended rod-like configuration might be expected, especially at low 
salt conditions. According to Tirado \al  \cite{Tirado84} the translational 
diffusion coefficient calculated for a rod-like macromolecule is given by
\begin{equation}
D^{th}=\frac{kT}{3\pi\eta}\frac{\ln(L_c/d)+0.312}{L_c}
\label{TiradoFormula}	
\end{equation}
Here $L_c=Nb$ is contour length, $d$ is polyion diameter, $\eta$ is viscosity of 
water  ($T= 298$ K).  Stellwagen \al \cite{Stellwagen03} have reviewed the 
literature and shown that the expression by Tirado \al~is well applicable to 
experimental data obtained for DNA molecules in size from 10 to 1000 basepairs. 
Then, the relationship which holds between the theoretical values should also 
hold for the experimental values obtained at varying DNA146 concentrations $c$.  
Thus, 
\begin{equation} D^{exp}_{146}(c)=\frac{D^{th}_{146}}{ D^{th}_{110*}} 
D^{exp}_{110*}(c) 
\label{extrapol}	
\end{equation} 
Using Eq.\ref{TiradoFormula} to get $D^{th}_{146}$ and $ D^{th}_{110*}$ and 
Eq.\ref{DDCy5} to get $D^{exp}_{110*}$ from the diffusion times $\tau$ measured 
for DNA110*, we directly convert $\tau$ into  $D^{exp}_{146}$. In this manner, 
fluorescence correlation spectroscopy provides the self-diffusion coefficient of 
DNA146 polyion, $D^{exp}_{146}(c)$ at varying concentrations ($c=0.0015-8$ mM, 
basepair). The results are shown in Fig.\ref{difftime}.

First, we note that $D^{exp}_{146}(c)$ is practically constant in 
the crossover concentration region $c=0.05-1$ mM identified from conductivity 
measurements (denoted by a shaded rectangle). $D^{exp}_{146}(c)$ only starts to 
vary above 1mM. This coincides with the dilute-semidilute crossover 
concentration for 50 nm long DNA146 molecules \cite{deGennes76}. At higher 
concentrations the polyions start to overlap and the apparent viscosity of the 
solutions changes, inducing the decrease of the diffusion coefficient 
\cite{Dhont}. The fact that our probe DNA110* "feels" the phenomenon (the 
dilute-semidilute crossover) due to DNA146 demonstrates that DNA110* diffusion 
properties indeed reflect the DNA146 diffusion.  

Second, below the crossover range $D^{exp}_{146}(c)$ starts to increase towards the 
value $D^{ss}$ (black triangle in Fig.\ref{difftime}) calculated, according to 
Eq.\ref{DDCy5} and Eq.\ref{extrapol} from $\tau^{ss}_{110*}$ obtained for the 
110 bases long ssDNA in pure water, without DNA146. This ssDNA is a sample of 
the Cy5 labeled synthetic oligonuclotide, dissolved in pure water, before any 
treatment (before mixing and hybridization with its complement). It is 
conceivable that $D^{ss}$ is the limiting value for a series of decreasing 
DNA146 concentrations. That is, due to the very low salt (practically without 
it: $c_{\mathrm{salt}}<0.05$mM) and rather low DNA concentration and 
correspondingly low counterion concentration \cite{Record75},  we presume that 
DNA denatures below 0.05 mM and becomes ssDNA.

\section{Discussion}
\label{sec4}

Osmometry for dsDNA \cite{Auer69,Raspaud00,Baigl05} has insofar been the primary 
experimental source of $\theta$ data of sufficient quality to validate 
extensions to Manning theory \cite{Hansen01}. Conductometry has been performed 
on different synthetic polymers, and has insofar given results for $\theta$ 
which only agree with Manning within a prefactor of the order of unity, and may 
depend strongly on the monomer concentration even in dilute solution 
\cite{Bordi04,Truzzolillo09,BordiJPC99,Wandrey99,Wandrey00}. We note that these 
experiments were either performed in semi-dilute solutions or with polydisperse 
samples and, most importantly, the synthetic polymers used were usually rather 
flexible. We remind that in these cases the conformation of the polyion is not 
well defined and renders analysis difficult due to the necessity to introduce a 
model for the conformation, besides the model for condensation and conductivity. 
However, modeling conformation of a flexible polyion in varying salt and monomer 
concentration is an elaborate problem in itself \cite{Bordi02,Dobrynin}. 

On the contrary, our DNA146 has well defined and simple rod-like conformation, 
it is highly monodisperse and forms a dilute solution. DNA in the very low salt 
conditions is also distinct as it is expected to go through melting transition 
with decreasing concentration, so we could have ssDNA or dsDNA in solution, 
depending on concentration \cite{Dove}. That is, this may allow us to compare 
$\theta(c)$ results to Manning values for both ssDNA and dsDNA in one 
experiment. Actually, counterion condensation is related to the DNA stability: 
entropic cost to condense or confine the counterions compares with the gain in 
electrostatic free energy upon DNA denaturation  \cite{Manning69,Manning72_a}. This gain 
is due to the single stranded DNA (ssDNA) having a lower linear charge density 
parameter than dsDNA, $u=1.7$ and $u=4.2$, respectively. Accordingly, the free 
counterions fraction should be higher for ssDNA, $\theta=0.59$ than for dsDNA, 
$\theta=0.24$. We have shown in the Introduction how such an increase in 
$\theta$ would lead to an increase in the polyelectrolyte conductivity, see 
Eq.\ref{IntroQuad}. 

Most 
importantly, we have measured the self-diffusion coefficient 
$D_p(c)=D^{exp}_{146}(c)$ of DNA146 as a function of DNA concentration. As we 
will show in the following, this allowed us to deconvolute the influence of DNA 
polyion charge, \ie~counterion condensation and the DNA polyion conformations on 
our conductometry data which is a function of both. We start with DNA polyion 
molar conductivity, defined by  

\begin{equation} 
\lambda_p=F 2\theta N {\mathrm e}\frac{D_p}{kT} 
\label{lambdap}	
\end{equation} 
First we note that this applies both for ssDNA and dsDNA. Comparing this 
expression with  Eq.\ref{FZDkT} and \ref{IntroQuad}, we find that for the 
valence we inserted $z_p=2$. This is due to  two negative charges (phosphate) 
being found on a single basepair in native dsDNA, which are still present on two 
separate nucleotides on two separated strands of ssDNA. Certainly, for an ssDNA 
of similar $N$ as an dsDNA $z_p$ equals one. However, since two ssDNA polyions 
appear in solution as a result of melting of one dsDNA molecule, the ssDNA 
concentration is doubled compared to dsDNA. This cancels the halved $z_p$, so 
there is no effect of melting on the polyelectrolyte conductivity $\sigma$, 
beyond the variation in $\theta$ or in $D_p$. Thus, for the sake of clarity, we 
can proceed by keeping the factor 2 within $\lambda_p$, nevermind the DNA state.

Inserting Eq.\ref{lambdap} into the expression for DNA conductivity 
Eq.\ref{normcondEq} we get (see also Eq.\ref{IntroQuad})

\begin{equation} 
\frac{\sigma(c)}{c}=2\theta\lambda_{{\mathrm {Na}}^+}+4\theta^2 ND_p(c)\frac{F\mathrm{e}}{kT} 
\label{beforeQuad}	
\end{equation}
This is a quadratic equation for $\theta(c)$ as a variable and 
$D_p(c)$ and $\sigma(c)$ as the parameters: 

\begin{equation} 
\theta(c)^2+\frac{\lambda_{{\mathrm {Na}}^+}}{D_p(c)\cdot cte.}\theta(c)-\frac{\sigma(c)}{c} \frac{1}{2D_p(c)\cdot cte.}=0 
\label{Quad}	
\end{equation} 
Here $cte.$  stands for a product of several constants (defined previously): 
$2NFe/kT$. Our measurements of the DNA146 polyelectrolyte conductivity 
$\sigma(c)$ and our independent probe of DNA146 diffusion coefficient 
$D_p(c)=D^{exp}_{146}$ ({\em cf.} Fig.\ref{normcond} and Fig.\ref{difftime} allow for the equation to be solved 
for the free counterion fraction $\theta$, without a necessity to model the DNA 
conformation. The equation is to be solved repeatedly for each concentration 
$c$, resulting in a concentration dependence $\theta(c)$. We take only the 
positive solutions as the physically meaningful.  

\begin{figure} 
\resizebox{0.46\textwidth}{!}{\includegraphics*{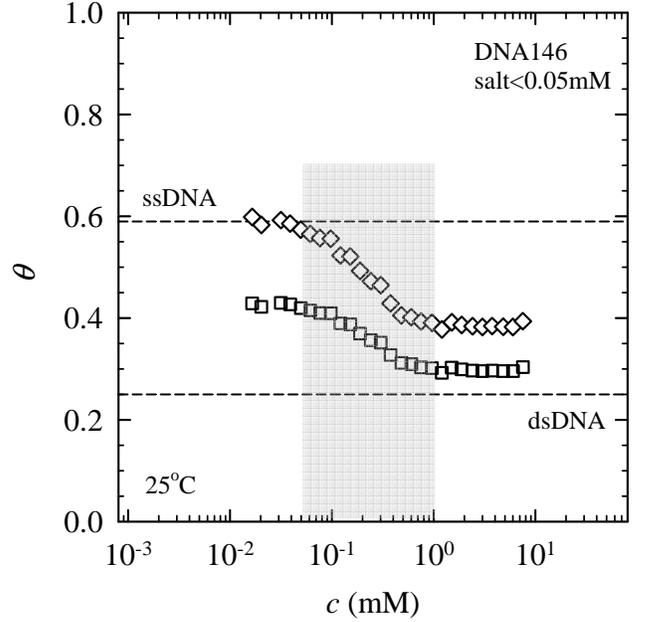}} 
\caption{ 
Free counterion fraction $\theta$  for 146bp nucleosomal DNA (DNA146) in pure 
water versus the DNA concentration $c$ (in mM basepairs). Squares denote $\theta 
(c) $ calculated according to  Eq.\ref{Quad}. Diamonds denote   $\theta'(c) $, 
the results of a calculation where the  assymetry field effect was taken into 
account, Eq.\ref{sigma_Manning_corr}. Dashed lines denote the theoretical value 
of $\theta=0.24$ for dsDNA and $\theta=0.59$ for ssDNA, as labeled. Shaded 
rectangle denotes the crossover concentration region identified from 
conductivity measurements.
} 
\label{fvsc} 
\end{figure}

The concentration dependence $\theta(c)$, according to Eq.\ref{Quad} for DNA146 
in pure water is shown (squares) in Fig.\ref{fvsc}. At lower concentrations it 
reaches a value $\theta=0.45$. Above 0.05 mM, in the conductivity crossover 
regime it starts to decrease, and above about 1 mM, outside crossover it becomes 
constant at $\theta=0.30$.  It is apparent that the experimentally derived range 
of values for $\theta$ falls within  the theoretical Manning values for ssDNA 
and dsDNA, as denoted by dashed lines in Fig.\ref{fvsc}.  Also, 
it may be noted that the 50\% variation in $\theta$ coincides with the conductivity 
crossover regime (denoted by the shaded rectangle). This is not surprising,  as 
$\sigma(c)/c$ is the only variable parameter in Eq.\ref{Quad}, while the polyion 
diffusion coefficient $D_p$ is rather constant in this regime.

The preceding calculation did not take into account the {\em asymmetry field} 
effect, due to the distortion of the counterion atmosphere surrounding the 
polyion, occurring when the polyion is subjected to an external electric field 
\cite{Manning81,Manning69,Wandrey99}. The original work by Manning  was reviewed 
and presented by Bordi \al\cite{Bordi04} in the form presented here. Asymmetry 
field effect can be taken \cite{Wandrey99} as if it corrects $\theta$ which 
appears in expressions for conductivity, ~Eq.\ref{sigmafzclambda} and 
Eq.\ref{normcondEq}, by a factor $B=0.866$. The factor, calculated by Manning, 
originates in the difference in the diffusion coefficients of counterions in the 
limit of infinite dilution and in the presence of polyions. Asymmetry field also 
influences the effective observable molar conductivity $\lambda_p$ of the 
polyion that figures in the above mentioned expressions. First,  it corrects the 
effective polyion charge through the factor $B$:

\begin{equation} 
\lambda_p'= 2 \theta B N D_p' \frac{F \mathrm{e}}{kT}
\label{lambdap_crtica}	
\end{equation} 
Second, the diffusion coefficient of the polyion is also corrected due to the asymmetry field: 
\begin{equation} 
D_p'=D_p\frac{1}{1+\frac{N D_p}{D_{{\mathrm {Na}}^+}} \cdot 2\theta B \cdot \frac{1-B}{B}} 
\label{Dp_crtica}	
\end{equation} 
Importantly, the diffusion coefficient of the polyion $D_p=D^{exp}_{146}(c)$ we 
have experimentally obtained by FCS, without electric field and thus without the 
asymmetry effect. Also, $D_{{\mathrm {Na}}^+}$ is the diffusion coefficient of 
free Na$^+$ ions in a simple, dilute electrolyte, $1.33 \cdot 10^{-9}$m$^2$/s. 
Combining the above, the Eq. \ref{normcondEq} becomes 
\begin{equation} 
\frac{\sigma}{c}= 2\theta B (\lambda_{{\mathrm {Na}}^+} + \lambda_p')
\label{sigmafzclambdaB}   
\end{equation} 
Inserting the Eqs. \ref{lambdap_crtica} and \ref{Dp_crtica} into Eq. 
\ref{sigmafzclambdaB} we get another equation which may be used to obtain the 
free counterion fraction: 

\begin{equation} 
\frac{\sigma}{c \cdot \lambda_{{\mathrm {Na}}^+}}=2\theta B \cdot (1 +2\theta B  \frac{N D_p}{D_{{\mathrm {Na}}^+}}\frac{1}{1+\frac{N D_p}{D_{{\mathrm {Na}}^+}} \cdot \frac{1-B}{B}\cdot 2\theta B}) 
\label{sigma_Manning_corr}	
\end{equation}

In analogy with Eq.\ref{Quad}, we rewrite this into a quadratic equation: 
\begin{equation} 
M/B x(c)^2+(1-AM \frac{1-B}{B})x(c)- A=0 
\label{Quad_2} 
\end{equation} 
$M$ stands for $\frac{N D_p}{D_{{\mathrm {Na}}^+}}$ and $A$ stands for 
$\frac{\sigma}{c\lambda_{{\mathrm {Na}}^+}}$. We remind that both $M$ and $A$ 
are obtained experimentally as functions of $c$, and that the equation was 
solved separately at each $c$ value, to get $x(c)$ (we only take the positive, 
physical solution). 

In Fig.\ref{fvsc} we show  the asymmetry field corrected $\theta'(c)=x(c)/2B$ 
(diamonds). Overall behaviour  of $\theta'(c)$ is analogous to $\theta(c)$ 
behaviour calculated with Eq.\ref{Quad}. The variation in $\theta(c)'$ also 
occurs within the conductivity crossover region (denoted by the shaded 
rectangle), and the relative change of  $\theta'$ in the crossover region 
remains about 50\%. However, the absolute values are different. At low 
concentrations $\theta'(c) =0.60$ reaches the theoretical value for ssDNA 
$\theta =0.59$, while at high concentrations it decreases only down to 
$\theta=0.4$.

The crossover in conductivity that we have observed for DNA in very low salt 
reflects as the crossover in $\theta(c)$ (or $\theta'(c)$), the Manning free 
counterion fraction. The exact $\theta$ values may depend whether basic 
corrections to polyelectrolyte conductivity are taken into account. 
Notwithstanding the details of the conductivity model \cite{Wandrey00}, we 
emphasize that the obtained extremal values for  $\theta$  correspond to Manning 
model predictions both for ssDNA and dsDNA (denoted in the Fig.\ref{fvsc} by 
dashed lines). This also corroborates the expected DNA melting across the 
studied DNA concentration range. 

Notably, our result complements the unique result  by Auer and Aleksandrowitz 
\cite{Auer69} obtained by osmometry for DNA solutions without added salt. These 
authors studied somewhat higher DNA concentration range 2-10 mM. For dsDNA they 
obtained an osmotic coefficient $\phi_0=0.16$ that would correspond to 
$\theta=0.32$ and for ssDNA they got $\phi_0=0.24$ corresponding to 
$\theta=0.48$ for ssDNA. The relationship between $\theta$ and $\phi_0$ is given 
by Manning \cite{Manning69,Wandrey00}. We find that it is very significant that 
both osmometry, and our technique find  $\theta$ for ssDNA only 
50\% larger than for dsDNA, while Manning condensation theory predicts more than 
100\%! While the  details of DNA conformations ({\em e.g.}~coiling or formation 
of hairpins in ssDNA) might be in the origin of this discrepancy, the limitations of Manning model 
should also be acknowledged - DNA is not a simple line charge.

The correspondence between the osmotic and transport measurements draws our 
final remark. That is, both techniques independently validate Mannings notion 
that counterions differentiate into two functionally separate populations.  
However, there is no {\em {a priori}} reason for these experiments to find 
similar fractions for these populations. The transport techniques measure the 
contribution to polyelectrolyte conductivity of the polyion whose charge is 
reduced due to the condensed counterions that move along, as well as the 
contribution of free counterions that move opposite to the polyion in the 
external electric field. Osmometry identifies as free the counterion fraction 
that contributes to the osmotic pressure of the solution. That is, those 
counterions that diffuse freely at distance from the polyion. However, as 
mentioned in the Introduction, the radial distribution of counterions is 
continuous and, beyond Manning model, in calculations based on Poisson-Boltzmann 
(PB) theory it is rather arbitrary to define any given distance from the polyion 
as the extent of condensed counterions zone  \cite{LeBretZimm84}.  Specifically, 
a nonlinear PB model has been worked out for a system very similar to our 
experimental one - rod-like polyelectrolyte dilute solutions in very low salt, 
and is based on defining the two (condensed and free) zones around the polyion 
\cite{Deshkovski01}. Now, according to our experiments, the condensed 
counterions zone radius should be less arbitrary. That is, as our conductometry 
study detects a reduced DNA polyion charge due to condensation and as the 
diffusion results indicate a DNA polyion diameter of 2.6 nm, then this is also 
the condensed counterions zone diameter. The condensed counterions are to be 
found in the immediate vicinity of the polyion, as initially suggested by 
Manning, and the cylindrical condensed counterions zone depicted in \cite{Deshkovski01} should be very thin.

\section{Summary and Conclusion} \label{sec5} 

In this work we have quantified Manning free (uncondensed) counterions fraction 
$\theta$ for dilute solutions of rod-like polyions - 150bp DNA fragments, in very 
low salt ($<0.05$ mM) and thus validated Manning condensation and conductivity 
theories devised for such a regime.

Our conductometry study revealed that the DNA solution molar conductivity 
normalized by DNA concentration, attains almost 100\% higher value below 0.05 mM  
than above 1 mM (basepair). The results for solutions of ssDNA (actually, 
samples of thermally denatured dsDNA) lacked this conductivity crossover. Then, 
we applied fluorescence correlation spectroscopy (FCS) to find that 
the diffusion coefficient of DNA polyion $D_p$ is practically constant in the crossover region. Thus, we have shown that the origin for 
the conductivity crossover lies in the increase of free charge fraction and 
decrease of the effective polyion charge, due to changes in Manning 
condensation, which we were able to quantify. Depending if the Manning 
asymmetry field effect was taken into conductivity model or not, we obtained the 
values within the ranges $\theta=0.40-0.60$ or  $\theta=0.30-0.45$, 
respectively.   

The conductivity crossover and $\theta$ variation are easily related to be due 
to DNA denaturation. However, the 50\% variation in  $\theta$ that we 
observe is smaller than what Manning condensation theory predicts as a 
difference between dsDNA and ssDNA (more than 100 \%). Nevertheless, a 50\% 
difference in $\theta$ between ssDNA and dsDNA was also obtained in osmotic 
pressure studies by other authors.  We also found surprising that variations in DNA 
conformation due to denaturation appear to be of lesser influence on the polyion 
conductivity.  The above two issues lead to the question how DNA conformations 
population changes with a decrease in DNA concentration in the very low salt 
environment. This is the subject of our following paper \cite{following}.

Further application of 
FCS, with samples subjected to an external electric field (similar as used for 
conductometry)  could quantitate asymmetry field effect on the diffusion 
coefficient of DNA146 polyion and reveal in detail to what extent the condensed 
counterions move with the polyion. Finally, combined conductometry and FCS 
studies of dilute monodisperse DNA in added salt solutions could extend the 
studies of $\theta$ and further complement the data obtained by osmometry.

\section*{Acknowledgement}

We gratefully acknowledge A.S. Smith and R. Podgornik for illuminating 
discussions. T.V. is thankful to S. Kempter for all her assistance in the lab. 
This work is based on the support from the Unity through Knowledge Fund, Croatia 
under Grant 22/08. The work was in part funded by IntElBioMat ESF activity. The 
group at the Institute of physics works within Project No. 035-0000000-2836 of 
Croatian Ministry of Science, Education and Sports.

\end{document}